\documentclass[dvipsnames,format=sigconf,nonacm]{acmart} %
\usepackage{graphicx} 
\usepackage[T1]{fontenc}
\usepackage{graphicx}
\usepackage{appendix}
\usepackage{mathtools}
\usepackage{color}
\usepackage{enumerate}
\usepackage{enumitem}
\usepackage{braket} 
\usepackage{hyperref}
\usepackage{amsmath}
\usepackage{amsfonts}
\usepackage{cleveref}

\newcommand{\blindrev}[1]{} 

\newcommand{\hri}{Honda Research Institute Europe GmbH, Carl-Legien Stra\ss e 30, 63073 Offenbach}
\newcommand{\aqa}{$\langle aQa ^L\rangle $ Applied Quantum Algorithms}
\newcommand{\liacs}{LIACS, Universiteit Leiden, Niels Bohrweg 1, 2333 CA Leiden}

\title{Comparing Qubit and Qudit Encodings for EV Charging and Trip Assignment Problems}

\author{Linus Ekstr\o m}
\affiliation{\hri\country{Germany}}
\author{Hao Wang}
\affiliation{\liacs, \\ \aqa\country{Netherlands}}
\author{Sebastian Schmitt}
\affiliation{\hri\country{Germany}}

\begin{document}

\begin{abstract}
Variational quantum algorithms have attracted attention for their potential to solve combinatorial optimization problems. 
We study how the choice of encoding affects the resource requirements and optimization behavior of a variational quantum optimization algorithm.
In order to quantify these effects, realistically inspired constrained electric vehicle (EV) fleet management problems were considered. 
These problems couple determining the optimal EV battery charging schedule with assigning EVs to trips requested by customers. 
We compare a conventional binary (qubit) trip encoding with an integer (qudit) encoding that represents assignments more directly. 
Both encodings guarantee the same feasible solution set, while the qudit encoding exponentially reduces the required Hilbert-space dimension.
We solve many random instances of highly constrained uni- and bi-directional charging problems using qudit-based  quantum approximate optimization algorithm (QAOA) and thoroughly evaluate the performance results. 
We find that the qudit encoding of customer trips achieves similar or better optimization performance at much reduced resource requirements and shorter simulation runtime.
These results highlight qudit-native encodings as a practical route for integer and multi-valued scheduling problems in variational quantum optimization.
\end{abstract}

\maketitle

\section{Introduction}
Variational quantum algorithms (VQAs)~\cite{tilly_VQE_review2022, cerezo2020variational,bharti2022noisy} are a leading approach for near-term quantum optimization, with the quantum approximate optimization algorithm (QAOA) as a central example~\cite{farhi2014quantum,blekos2024QAOA_Review,bharti2022noisy}. 
In practical applications, the encoding of decision variables into quantum degrees of freedom can strongly affect both resource requirements and the resulting variational landscape~\cite{Sawaya2020encoding,Sawaya2023encodingtradeoffs,ChenDomainWall2021,karimiInegerEncoding2019,matteoEncoding2021}. 

Recently, high-dimensional qudits~\cite{Wang2020} have emerged as a promising technological paradigm for quantum computation.
Qudit technologies have shown benefits in a range of topics from algorithmic improvements to lattice gauge field theory simulation and quantum error correction~\cite{xiaoyang2026efficientquditalgo,meth2025latticegaugequdit,li2026transversalquantumcodes,joshi2025efficientquditcircuitquench}.
The benefits are especially relevant for optimization problems with integer variables, where qudit-based formulations can represent decisions directly rather than via binary expansions. 
Recent work has developed qudit QAOA for integer optimization problems~\cite{deller2023quantum,ekstrom2025qmoo}, alongside approaches for efficient handling of constraints in mixed-dimensional circuits~\cite{bottarelliConstraints2024}.

We study a realistic problem of operating a fleet of electric vehicles (EVs), which amounts to scheduling the bi-directional charging of each EV, and at the same time, serving trips requested by customers.  
Joint trip and charging scheduling has been widely studied in the operations research and energy systems literature because it combines assignment decisions, time coupling, and operational constraints such as battery state-of-charge (SOC) limits, trip overlaps, and grid power bounds~\cite{mendek2024case,limmer_EV_LNS_2023,singh_limmer_EVcharging_MOO2022,vargaLimmerEVCharging2022}. 
This structure makes the problem a useful testbed for comparing quantum encodings under realistic constraint patterns.

We compare two representations of the trip-assignment variables. The binary encoding uses one qubit variable per EV--trip pair. 
The integer encoding uses one qudit variable per trip whose value directly specifies the assigned EV, or an unserved option. 
In both cases, charging decisions are represented by qudit variables. 
This isolates the effect of the trip encoding on Hilbert-space dimension and optimization behavior.

Using exact state-vector simulations with shot-based cost estimation, we benchmark both encodings on small instances of the full problem. 
We compare Hilbert-space dimension, feasibility, and success statistics, and runtime. We find that the qudit trip encoding yields a substantially smaller search space and lower runtime, while achieving slightly better optimization performance in the tested regime. 
This supports qudit-native encodings as a practical choice for constrained scheduling problems with discrete multi-valued variables.

The paper is organized as follows. 
\Cref{sec:problem} introduces the EV charging and trip assignment problem. 
\Cref{sec:quantum} presents the quantum formulation and QAOA setup. 
\Cref{sec:benchmark-setup} explains the benchmark problems used in our comparison.
\Cref{sec:results} reports benchmark results and compares and discusses the results of the encodings. 
Finally, in \Cref{sec:conclusion} we discuss the implications of this study and lay down some thoughts about future directions of inquiry.

\section{Bi-directional EV Charging and Trip Assignment}
\label{sec:problem}
Efficiently operating a fleet of electric vehicles (EVs) is increasingly important as electro-mobility services become more and more popular.
In that context, bi-directional charging is very promising as vehicles are not used continuously but are parked for substantial time periods during the day. 
Therefore, the batteries of the EVs represent a resource that could be utilized while EVs are not driving~\cite{mendek2024case}. 
Technologies such as vehicle-to-grid (V2G) and vehicle-to-home (V2H) enable fleet operators to use idle vehicles as active energy assets rather than passive loads.
EV fleet operators offer transportation services to their customers, where the challenge is to allocate EVs to customer requests (trips) while charging the EVs during time periods where they do not serve trip requests.
This leads to a constrained optimization problem that couples trip assignment and charging decisions.

We consider a fleet of $N_{EV}$ EVs, over a horizon of $T$ discrete time steps.
Each EV $n$ has initial battery SOC $E_n^{0}$ at $t=0$ and must satisfy a minimum final SOC $E_n^{\min}$ at $t=T-1$. 
At each time step $t$, EV $n$ chooses a discrete charging power level $l_{n,t}$, where positive values denote charging, negative values denote discharging, and zero denotes idling. 
The electricity price at time step $t$ is $c_t$, and grid (fuse) bounds constrain the total power $\sum_n l_{n,t}$. 
The system receives $R$ trip requests. Each trip $i$ has a time window $T_i=[t_i^{\mathrm{start}},t_i^{\mathrm{end}}]$ and energy demand $E_i^{\mathrm{trip}}$. 
Trips are assigned to at most one EV; overlapping trips cannot share an EV, and EVs cannot charge or discharge while serving a trip. 
We seek a charging schedule giving minimum cost while fulfilling all the technical constraints described in \Cref{sec:trip_encodings}.

The charging variables are $d$-ary integers with $d=2l_{\max}+1$, i.e.\  $l_{n,t}\in\{-\tfrac{d-1}{2},\dots,0,\dots,\tfrac{d-1}{2}\}$ ($d$ odd) for bi-directional charging and  $d=l_{\max}+1$, i.e.\ $l_{n,t}\in\{0,\dots,d-1\}$, for uni-directional charging. 
For the trip-assignment variables, we compare binary and integer encodings.

\subsection{Trip-assignment encodings}
\label{sec:trip_encodings}
In the binary encoding, we introduce one binary variable $r_{n,i}\in\{0,1\}$ for each EV--trip pair, where $r_{n,i}=1$ means that EV $n$ serves trip $i$. The objective is
\begin{align}
\label{eq:cost_binary}
C_{\mathrm{bin}}(\{l_{n,t},r_{n,i}\})
&=
\Delta t \sum_{t=1}^{T}\sum_{n=1}^{N_{EV}} c_t\, l_{n,t}
+ \lambda\sum_{i=1}^{R}\delta_{0,\sum_{n=1}^{N_{EV}} r_{n,i}},
\end{align}
where $\delta_{ij}$ is the Kronecker delta function. 
The first term of \Cref{eq:cost_binary} is the charging cost. 
The second term increases the cost associated with unserved trips by adding $\lambda$ whenever $\sum_n r_{n,i}=0$. $\lambda$ also sets the trade-off between charging cost and trip fulfillment.

The binary constraints are
\begin{align}
&\sum_{n\neq m} r_{n,i} r_{m,i}=0 \hspace{15pt} \forall i
\label{eq:bin_one_ev}\\
& r_{n,i} r_{n,j}=0 \hspace{32pt} \forall n,\; i\neq j \quad \text{if } T_i\cap T_j\neq \emptyset
\label{eq:bin_no_overlap}\\
& r_{n,i} l_{n,t}=0 \hspace{34pt} \forall n,i,\; t\in T_i
\label{eq:bin_no_charge}\\
& E_n^0+\Delta t\sum_{k<t} l_{n,k} - \sum_{t_i^{\mathrm{start}}\le t} E_i^{\mathrm{trip}} r_{n,i}\ge 0 \hspace{26pt} \forall n,t
\label{eq:bin_soc_low}\\
& E_n^0+\Delta t\sum_{k<t} l_{n,k} - \sum_{t_i^{\mathrm{start}}\le t} E_i^{\mathrm{trip}} r_{n,i}\le E_n^{\mathrm{cap}} \hspace{16pt} \forall n,t
\label{eq:bin_soc_up}\\
& E_n^0+\Delta t\sum_t l_{n,t} - \sum_i E_i^{\mathrm{trip}} r_{n,i}\ge E_n^{\mathrm{min}} \hspace{25pt} \forall n
\label{eq:bin_soc_final}\\
& P^{\min}\le \sum_n l_{n,t}\le P^{\max} \hspace{15pt} \forall t
\label{eq:bin_grid}
\end{align}
with $r_{n,i}\in\{0,1\}$, $l_{n,t}\in\left\{-\tfrac{d-1}{2},\dots,0,\dots,\tfrac{d-1}{2}\right\}$   for bi-directional charging, and $l_{n,t}\in\left\{0,\dots,d-1\right\}$ for uni-directional charging. 
\Cref{eq:bin_one_ev,eq:bin_no_overlap} enforce valid assignments, i.e.\ that each trip is served by at most one EV, and each EV cannot serve two trips whose time windows overlap. 
\Cref{eq:bin_no_charge} prevents charging or discharging while an EV is assigned to a trip. 
\Cref{eq:bin_soc_low,eq:bin_soc_up} keep the SOC within $[0,E_n^{\mathrm{cap}}]$ at all intermediate times, and \Cref{eq:bin_soc_final} enforces the required final SOC. 
Finally, \Cref{eq:bin_grid} enforces grid power bounds, set by $P^{\text{min}}$ and $P^{\text{max}}$, at each time step.

In the integer encoding, we introduce one integer variable $q_i\in\{0,\dots,N_{EV}\}$ per trip. 
If $q_i=n$, trip $i$ is served by EV $n$, while $q_i=0$ denotes an unserved trip. 
The objective is
\begin{align}
\label{eq:cost_int}
C_{\mathrm{int}}(\{l_{n,t},q_i\})
&=
\Delta t \sum_{t=1}^{T}\sum_{n=1}^{N_{EV}} c_t\, l_{n,t}
+ \lambda\sum_{i=1}^{R}\delta_{q_i,0} .
\end{align}
The objective in \Cref{eq:cost_int} has the same structure as in the binary case. 
Charging cost is linear in $l_{n,t}$, and unserved trips are penalized by $\delta_{q_i,0}$.

The integer constraints are
\begin{align}
& q_i\neq q_j \hspace{36pt} \forall i\neq j \quad \text{if } T_i\cap T_j\neq \emptyset
\label{eq:int_no_overlap}\\
& \delta_{q_i,n} l_{n,t}=0 \hspace{20pt} \forall n,i,\; t\in T_i
\label{eq:int_no_charge}\\
& E_n^0+\Delta t\sum_{k<t} l_{n,k} - \sum_{t_i^{\mathrm{start}}\le t} E_i^{\mathrm{trip}} \delta_{q_i,n}\ge 0 \hspace{26pt} \forall n,t
\label{eq:int_soc_low}\\
& E_n^0+\Delta t\sum_{k<t} l_{n,k} - \sum_{t_i^{\mathrm{start}}\le t} E_i^{\mathrm{trip}} \delta_{q_i,n}\le E_n^{\mathrm{cap}} \hspace{15pt} \forall n,t
\label{eq:int_soc_up}\\
& E_n^0+\Delta t\sum_t l_{n,t} - \sum_i E_i^{\mathrm{trip}} \delta_{q_i,n}\ge E_n^{\mathrm{min}} \hspace{24pt} \forall n
\label{eq:int_soc_final}\\
& P^{\min}\le \sum_n l_{n,t}\le P^{\max} \hspace{15pt} \forall t
\label{eq:int_grid}
\end{align}
where $q_i\in\{0,\dots,N_{EV}\}$ and $l_{n,t}$ the same as for the qubit encoding. 
The constraint set mirrors the binary formulation. \Cref{eq:int_no_overlap} prevents assigning overlapping trips to the same EV, because equal values of $q_i$ and $q_j$ would indicate the same vehicle. 
\Cref{eq:int_no_charge} forbids charging or discharging during the trip window of an assigned EV. \Cref{eq:int_soc_low,eq:int_soc_up} enforce SOC bounds at all times, and \Cref{eq:int_soc_final} enforces the minimum final SOC. \Cref{eq:int_grid} enforces the grid bounds.

In both encodings, trip energy is subtracted at the trip start time $t_i^{\mathrm{start}}$. 
The integer encoding removes the explicit one-EV-per-trip constraint, since that condition is automatically guaranteed by the encoding.

The total number of configurations is
\begin{align}
N_{\mathrm{bin}}&=d^{N_{EV}T}2^{N_{EV}R},\\
N_{\mathrm{int}}&=d^{N_{EV}T}(N_{EV}+1)^R,
\end{align}
which gives
\begin{align}
\frac{N_{\mathrm{int}}}{N_{\mathrm{bin}}}
=2^{-RN_{EV}+R\log_2(N_{EV}+1)}\,
\end{align}
as the ratio of the two encodings.
The number of configurations is exponentially reduced by the integer encoding compared to the binary encoding as a function of the number of trips $R$, and the number of EVs $N_{EV}$ (with logarithmic corrections in the latter case).

\section{Quantum Formulation}
\label{sec:quantum}
We compare the two trip-encoding variants of \Cref{sec:problem} in a QAOA-based quantum optimization framework, with the goal of isolating how the trip encoding affects optimization performance and resource requirements.

To obtain the quantum formulation, we choose a computational basis and replace the classical decision variables with the corresponding quantum operators. 
All constraints are incorporated through non-linear penalty terms in the cost Hamiltonian. 
In the exact state-vector simulations used here, these non-linear terms can be implemented directly in the computational basis.

\subsection{Quantum Representation of the Trip Encodings}
\label{sec:quantum_trip_encodings}
We use the same qudit charging variables $l_{n,t}$ in both formulations. The difference is how the trip assignment is represented in the computational basis. 
For the binary encoding, the basis states are $\ket{l_{n,t},r_{n,i}}$, where $r_{n,i}\in\{0,1\}$ indicates whether EV $n$ serves trip $i$. 
The diagonal operators act in the computational basis as
\begin{align}
L^z_{n,t}\ket{l_{n,t},r_{n,i}} &= l_{n,t}\ket{l_{n,t},r_{n,i}},
\\
R^z_{n,i}\ket{l_{n,t},r_{n,i}} &= r_{n,i}\ket{l_{n,t},r_{n,i}}
.
\end{align}
Meanwhile, for the integer encoding, the basis states are $\ket{l_{n,t},q_i}$, where $q_i\in\{0,\dots,N_{EV}\}$ specifies the assigned EV and $q_i=0$ denotes a trip not served by any EV. 
The diagonal operators act in this computational basis as
\begin{align}
L^z_{n,t}\ket{l_{n,t},q_i} &= l_{n,t}\ket{l_{n,t},q_i},
\\
Q^z_{i}\ket{l_{n,t},q_i} &= q_i\ket{l_{n,t},q_i}
.
\end{align}
In both cases, we use the standard angular-momentum operators, with $L^{x,y}_{n,t}$, $R^{x,y}_{n,i}$, and $Q^{x,y}_{i}$ defined in the usual way from ladder operators~\cite{deller2023quantum}. 
More details can be found in \Cref{sec:ang-momentum}.  

The resources required by each encoding directly follow the number of configurations given in~\Cref{sec:trip_encodings}.
The binary encoding uses $N_{EV}R$ qubits and $N_{EV}T$ qudits of dimension $d$, which gives the Hilbert space dimension
\begin{align}
\dim(\mathcal{H}_{\mathrm{bin}})=d^{N_{EV}T}\,2^{N_{EV}R}=N_{\mathrm{bin}}.
\end{align}
The integer encoding uses $R$ qudits of dimension $N_{EV}+1$ and $N_{EV}T$ qudits of dimension $d$, which yields
\begin{align}
\dim(\mathcal{H}_{\mathrm{int}})=d^{N_{EV}T}\,(N_{EV}+1)^R = N_{\mathrm{int}}.
\end{align}
The Hilbert space dimension of the qudit encoding is again exponentially reduced in $R$ and $N_{EV}$ compared to the qubit encoding,
\begin{align}
\label{eq:hilbertDim}
\dim(\mathcal{H}_{\mathrm{int}})
= \dim(\mathcal{H}_{\mathrm{bin}})\,2^{-RN_{EV}+R\log_2(N_{EV}+1)}.
\end{align}
This shows that the qudit encoding is much more efficient in the required quantum resources than the traditional qubit encoding.

\subsection{Quantum Optimization Algorithm}
\label{sec:quantum-optimization-algorithm}
We construct the cost Hamiltonian by promoting the classical variables $l_{n,t}$, $r_{n,i}$, and $q_i$, in \Cref{eq:cost_binary,eq:cost_int}, to the corresponding diagonal operators $L^z_{n,t}$, $R^z_{n,i}$, and $Q^z_i$. 
All constraints are enforced via penalty terms added to the cost Hamiltonian,

\begin{align}
    \label{eq:ev_ham_penalty}
    H_C=&\: C(\{l_{n,t}\to L^z_{n,t} , r_{n,i}\to R_{n,i}^z, q_i\to Q_{i}^z\}) \\ \nonumber
        & + \alpha 
            \sum_e E_e(\{L^z_{n,t} , R_{n,i}^z, Q_{i}^z\}) 
        + \alpha \sum_f F_f(\{L^z_{n,t} , R_{n,i}^z, Q_{i}^z\})
\end{align}
where $\alpha>0$ is a penalty factor. 
The functions $E_e$ and $F_f$ encode equality and inequality constraints, respectively, and contribute only when constraints are violated. 
For example, the no-charging-during-trip constraint \Cref{eq:bin_no_charge} can be enforced by
\begin{align}
    E_1 & =\sum_{n,i}\sum_{t\in T_i} \big|R^z_{n,i}L^z_{n,t}\big|\\
    F_1 & =\sum_t \big|P^{\min}-\sum_n L^z_{n,t}\big|\Theta\Big(P^{\min}-\sum_n L^z_{n,t}\Big)
\end{align}
with $\Theta(x)$ the Heaviside step function and $|x|$ the absolute value (not to be confused with an operator norm). 
These non-linear functions of operators are shorthand notations and can be viewed as being defined by the series expansions.
Practically, these functions transform to functions of the eigenvalues when applied to computational basis states, e.g.,\ $f(L^z_{n,t})\ket{l_{n,t}}=f(l_{n,t})\ket{l_{n,t}}$.
In hardware-oriented implementations, such nonlinear penalties require additional techniques \cite{yarkoni2021,bottarelliConstraints2024,djidjev_Lagrangian_constrains_2023,kuroiwa_PenalyVQE_2021}.
These techniques incur additional overheads in terms of qubits and qudits, as well as additional multi-qubit/qudit interactions, i.e., \ two-body gates.    
However, here we use exact state-vector simulations and implement the penalty terms directly.
We focus on encoding-induced structural differences rather than on overheads related to hardware implementations.

We employ an $L$-layer QAOA circuit of the form
\begin{align}
\label{eq:ev_Ucirc}
    U(\boldsymbol{\gamma},\boldsymbol{\beta})= \prod_{l=1}^{L}e^{-i {H}_X(\beta_l)}  e^{-i\gamma_l H_C},
\end{align}
where $\boldsymbol{\gamma}=(\gamma_1,\dots,\gamma_L)$ and $\boldsymbol{\beta}=(\beta_1,\dots,\beta_L)$ are the variational parameters for the cost and mixing unitaries, respectively. 
We use the qudit mixer of~\cite{bottarelliConstraints2024,deller2023quantum,ekstrom2025qmoo},
\begin{align}
\label{eq:ev_mix}
H_X(\beta_l)=\beta_{1;l}\sum_i L_i^x+\beta_{2;l}\sum_i (L_i^z)^2 ,
\end{align}
which introduces two mixing parameters per layer.

Given an easy to prepare initial state $\ket{\psi_0}$, the circuit prepares the final state $\ket{\psi(\boldsymbol{\gamma},\boldsymbol{\beta})}=
U(\boldsymbol{\gamma},\boldsymbol{\beta})\ket{\psi_0}$. 
The classical objective function to be minimized is given by 
\begin{align}
\label{eq:costTrace}
C(\boldsymbol{\gamma},\boldsymbol{\beta})
&= \braket{\psi(\boldsymbol{\gamma},\boldsymbol{\beta}) | H_C | \psi(\boldsymbol{\gamma},\boldsymbol{\beta})}\,.
\end{align}
Despite performing essentially exact state-vector simulations in this study, we already anticipate realistic implementation on hardware where expectation values like \Cref{eq:costTrace} cannot be calculated exactly.
Instead, one needs to revert to approximate estimations using a finite number of  measurement shots 
\begin{align}
\label{eq:costTraceShot}
C(\boldsymbol{\gamma},\boldsymbol{\beta}) 
\approx 
& \frac1M \sum_{\substack{m=1\\z_m \sim p_{\boldsymbol{\gamma},\boldsymbol{\beta}}(z)}}^M \braket{z_m  | H_C | z_m}
=\frac1M \sum_{\substack{m=1\\z_m \sim p_{\boldsymbol{\gamma},\boldsymbol{\beta}}(z)}}^M E_C(z_m) 
\,,
\end{align}
where $z_m$ indicates configurations defined by computational basis states, $\ket{z}=\ket{\{l_{n,t},r_{n,i}\}}$ for the qubit encoding, and 
$\ket{z}=\ket{\{l_{n,t},q_{i}\}}$ for the qudit encoding. 
The configurations are sampled following the Born-rule from the probability distribution defined by the quantum state, $p_{\boldsymbol{\gamma},\boldsymbol{\beta}}(z)=|\braket{z|\psi(\boldsymbol{\gamma},\boldsymbol{\beta})}|^2$.
$\,E_C(z)$ denotes the classical cost function including penalty terms as indicated by \Cref{eq:ev_ham_penalty}.

Our implementation of QAOA uses the equal superposition of all basis states as initial state, $\ket{\psi_0}=1/\sqrt{\text{dim}(\mathcal{H})}\sum_z\ket{z}$, and employs a classical optimization loop which minimizes the cost function of \Cref{eq:costTraceShot} by optimizing the variational parameters \(\boldsymbol{\gamma}\) and \(\boldsymbol{\beta}\). 
The dimensionality of the search space for this optimization is determined by the number of QAOA layers as $3L$.
For systems with only qubits, the second mixing term $\sim (L^z)^2$ is proportional to the identity, and therefore only adds a global phase, which is irrelevant for the optimization, so the number of parameters is reduced to $2L$ in this case.

\section{Benchmark Setup}
\label{sec:benchmark-setup}
We benchmark the two encodings on randomly generated instances from two classes of problems.

The first class of problems considers bi-directional charging with three charging levels ($d=3$), using $N_{EV}=3$ EVs, $T=2$ time steps, and $R=2$ trip requests. 
This yields Hilbert-space dimensions $\dim(\mathcal{H}_{\mathrm{bin}})=3^{6}\cdot2^{6}=46656$ and $\dim(\mathcal{H}_{\mathrm{int}})=3^{6}\cdot4^{2}=11664$ for the qubit and qudit trip encodings respectively. 

The second class considers unidirectional charging problems with only two charging levels ($d=2$) corresponding to idling and charging an EV. 
For this problem class, we use $N_{EV}=2$ EVs, $T=3$ time steps, and vary the number of trip reservations from $R=2$ to $R=4$ to investigate the impact of the trip encoding.
The Hilbert-space dimensions are $\dim(\mathcal{H}_{\mathrm{bin}})=2^{6}\cdot2^{2R}=\{1024,4096,16384\}$ and $\dim(\mathcal{H}_{\mathrm{int}})=2^{6}\cdot3^{R}=\{576, 1728, 5184\}$ for qubit and qudit trip encoding. 
While the Hilbert space dimension differs between encodings, the number of feasible states remains identical by construction. 
Consequently, cost values coincide on feasible configurations and differ only on infeasible ones (due to encoding-dependent penalties). 

For each problem type we generate ten problem instances by uniformly sampling in the given parameter ranges as follows: charging cost values $c_t\sim\mathcal{U}(0.3,4)$, trip start and end times $t_i^{\mathrm{start}}=t_i^{\mathrm{end}}\sim\mathcal{U}(\{0,T-1\})$, required electrical energy for each EV $E_n^{\min}-E_n^{0}\sim\mathcal{U}(0,2)$, and electricity cost for each trip $E_i^{\mathrm{trip}}\sim\mathcal{U}(0,2)$. 
We set $P^{\max}=-P^{\min}=10$, so grid bounds are inactive for these instances. 
We assumed dimensionless variables, i.e., \ set $\Delta t=1$ and $\lambda=3$, and used a penalty factor of  $\alpha=10$.

\begin{figure}
     \centering
     \includegraphics[width=0.47\linewidth]{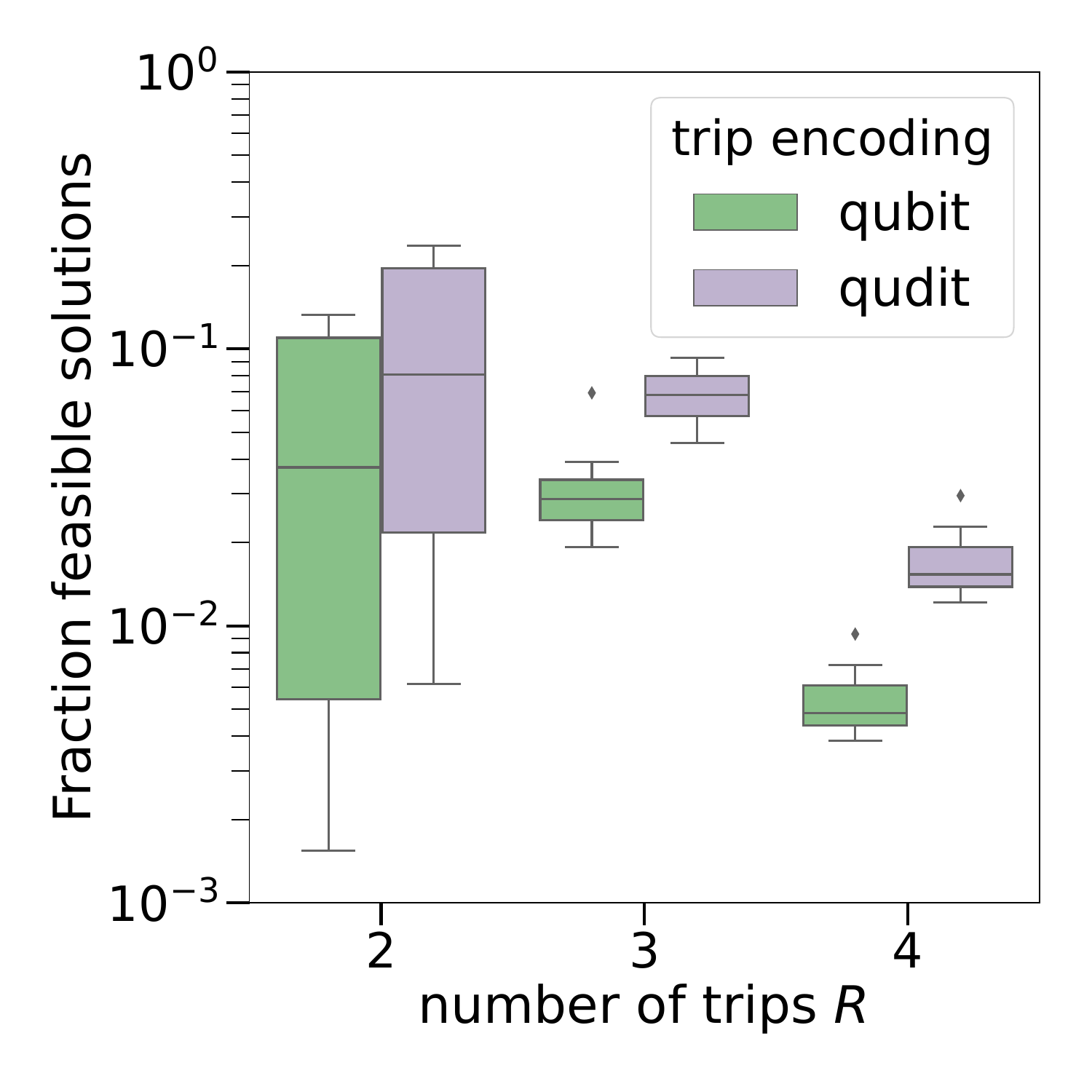}
     \includegraphics[width=0.47\linewidth]{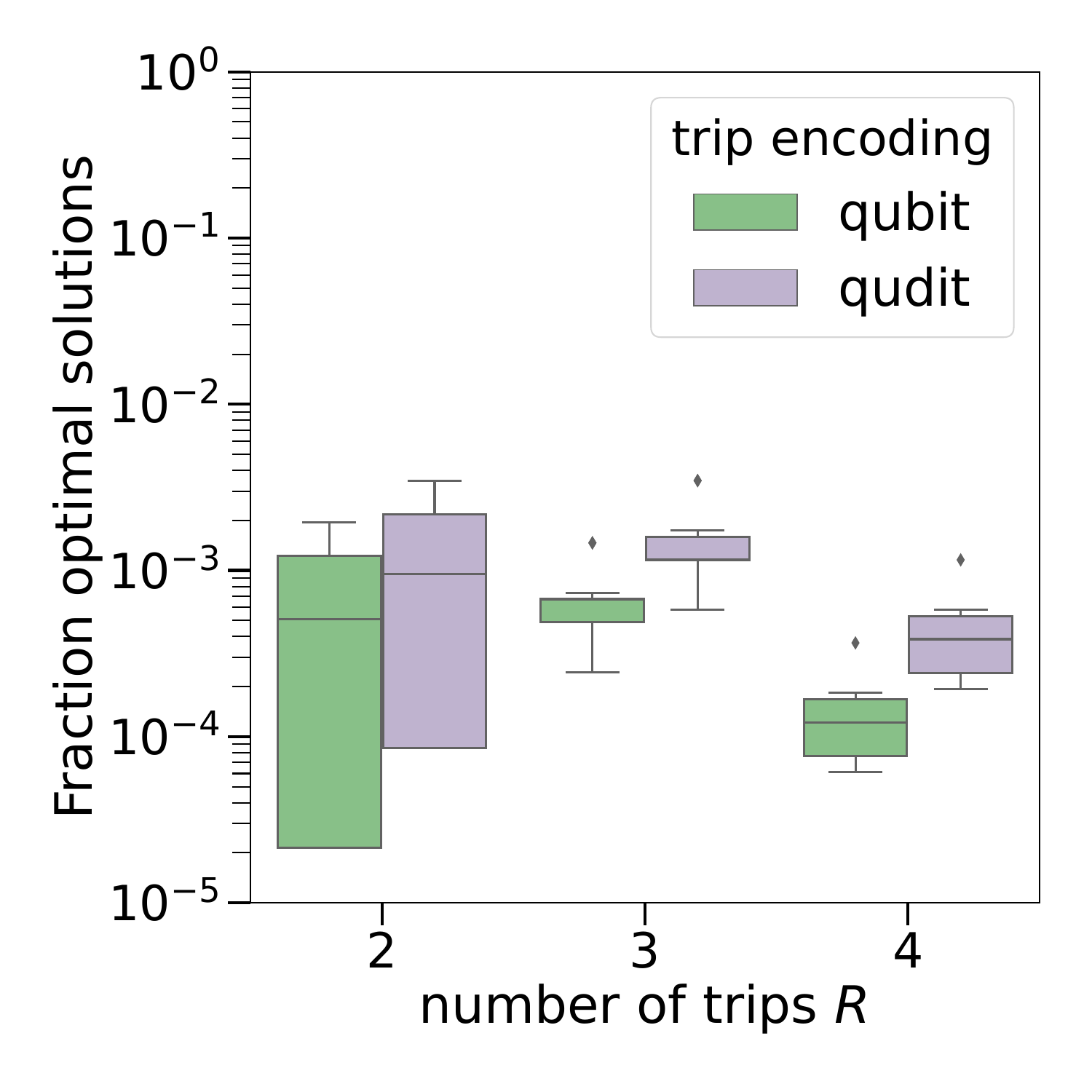}
     \caption{Statistics of all the benchmark problem instances as a function of the number of trips for the two encodings. 
     Left: relative number of feasible solutions. Right: relative number of optimal solutions.
         The boxes represent the inter-quartile ranges, the whiskers show the extent of the full distribution, the points indicate outliers, and the horizontal lines inside the boxes are median values. 
         The problems considered in this work are highly constrained with low percentages of feasible and optimal solutions which decrease with increasing number of trip requests $R$. 
        }
\label{fig:problem_stats}
\end{figure}

The resulting problem instances are highly constrained as illustrated in \Cref{fig:problem_stats}. 
The statistics of the fraction of feasible configurations of the problems are shown in the left panel, and reveal that only about $0.1\%$  to $20\%$  of all configurations are feasible. The number of feasible solutions decreases as the number of trips, $R$, increases. 
It is clear from \Cref{eq:hilbertDim} that the qubit encoding has lower fractions of feasible states. 
The number of optimal solutions is very small, as visible in the right panel of \Cref{fig:problem_stats}.

We employ the Powell optimizer from the \texttt{scipy} library~\cite{2020SciPy-NMeth} for optimizing the variational parameters of the quantum circuits.  
For each problem instance, we perform ten full QAOA runs, each time using initial parameters randomly sampled from a uniform distribution on $[0,2\pi]$.
We fix the number of internal iterations for the Powell optimizer to be 200. 
To estimate the cost function values in \Cref{eq:costTrace} during optimization, we always use $M=256$ measurement shots, yielding a total shot budget of $51200$ measurements per full optimization run.

\section{Quantum Optimization Results}
\label{sec:results}
We report benchmark results for the full EV charging and trip assignment problem, comparing the two trip-encoding variants, and present optimization run results using several metrics.

One central performance criterion for the optimization is the mean energy gap, defined as
\begin{align}
\label{eq:ev_Emean}
\Delta E_{mean} = \frac{1}{M}\sum_{m=1}^{M} E_C(z_m) - E_{GS}\,,
\end{align}
where $z_m$ are the $M$ measurement samples obtained from the optimized circuit (see \Cref{eq:costTraceShot}), $E_C(z)$ is the cost function estimation from the output of the quantum circuit (including penalties), and $E_{GS}$ is the exact optimum obtained by full enumeration. 

A related performance criterion is the minimal energy gap, defined as
\begin{align}
\label{eq:ev_Emin}
\Delta E_{min} = \min_m E_C(z_m) - E_{GS}\,,
\end{align}
which represents the best solutions sampled from the final optimized quantum circuit.

Another important metric is whether the global optimum is found during optimization. 
This can be summarized in the  success probability,  
\begin{equation}
\label{eq:ev_success}
    r = \frac{1}{N_{\mathrm{runs}}}\sum_{i=1}^{{N}_{\mathrm{runs}}}X_i\,,
\end{equation}
where $N_{\mathrm{runs}}=10$ is the total number of runs and  $X_i=1$ if the algorithm sampled at least one of the (possibly degenerate) optimal states in the $M$ samples of the optimized quantum circuit of the $i-$th run, and $X_i=0$ otherwise.  

Similarly, the fraction of feasible states sampled from the final optimized quantum circuit also provides valuable information, 
\begin{equation}
\label{eq:ev_feasible}
    f =  \frac{1}{N_{\mathrm{runs}}}\sum_{i=1}^{{N}_{\mathrm{runs}}}\frac{F_i}{M}\,,
\end{equation}
where $F_i$ is the number of feasible states obtained from sampling $M$ states from the optimized quantum circuit of the $i-$th run.

In the analysis section below, we report performance metrics averaged over all problem instances.

\subsection{Bi-directional charging}
\label{sec:bi-direcionalResults}

\begin{figure*}
    \centering
    \hspace*{-1cm}qubit trip encoding\hspace*{0.32\linewidth} qudit trip encoding\\
    \includegraphics[width=0.45\linewidth,trim={10mm 5mm 2mm 10mm},clip]{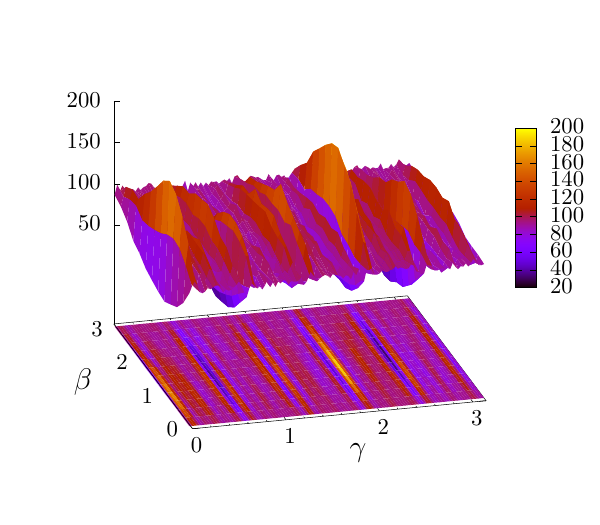} 
    \includegraphics[width=0.45\linewidth,trim={10mm 5mm 2mm 10mm},clip]{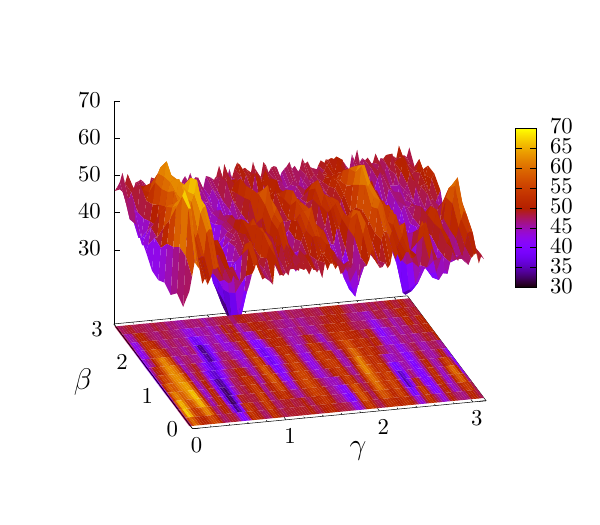} 
    \caption{Cost function landscape for a $L=1$ layer QAOA as function of the variational parameters $\beta$ and $\gamma$ for one example of the bi-directional ($d=3$) benchmark problem instance with $N_{EV}=3$, $T=2$ and $R=2$. Left: qubit trip encoding. Right: qudit trip encoding. We only include the $L^x$ mixer and not the $(L^z)^2$ term of \Cref{eq:ev_mix} (i.e.\ we set $\beta_2=0$). 
    The cost functions are evaluated with $M=256$ shots. 
    The cost function landscapes are highly multi-modal with very different  oscillation frequencies  for the mixing  ($\beta$) and phase  ($\gamma$) parameters. The overall scale for qudit encoding is lower than for the qubit encoding. 
    }
    \label{fig:costlandscape}
\end{figure*}

\Cref{fig:costlandscape} shows a typical cost landscape as a function of the two variational parameters for a one-layer ($L=1$) QAOA.
In this example, we show an instance of the bi-directional charging problem with $N_{EV}=3$, $T=2$, and $R=2$. 
We set the second mixing parameter $\beta_2=0$ in order to be able to plot a two-dimensional cost landscape.
The left panels in \Cref{fig:costlandscape} show the landscape for the binary qubit encoding of the requested trips, while the right panel show the exact same problem instance but for integer qudit trip encoding.

First, it can  be observed that both landscapes are highly multi-modal and show heavy oscillatory behavior. 
The overall scale for the variation with the mixing parameter $\beta$  is of the order of unity, which is understandable as the corresponding unitary operator induces factors of $e^{i \beta \phi_X}$ with $\phi_X\sim 1$ of the order of one.  
In contrast, the scale of the cost unitary is determined by the cost function values of basis states, which are much larger than unity, and consequently lead to much higher oscillation frequencies, $e^{i \gamma \phi_C}$ with $\phi_C\sim10-100$.
This asymmetry between the characteristic behavior for mixing ($\beta$) and cost parameters ($\gamma$) makes the optimization problem even harder.
Additionally, the finite number of $M=256$  measurement shots used to evaluate cost function expectation values leads to shot noise which can be clearly seen in the figure as well. 
Comparing the landscapes, one can observe that the overall energy scale for the qubit encoding is about a factor $\sim3$  larger than the qudit encoding. 
This is due to the much larger fraction of infeasible states in the qubit encoding, which typically have much larger energies because of contributions from the penalty terms.  

\begin{figure*}
    \centering
    \includegraphics[width=0.24\linewidth]{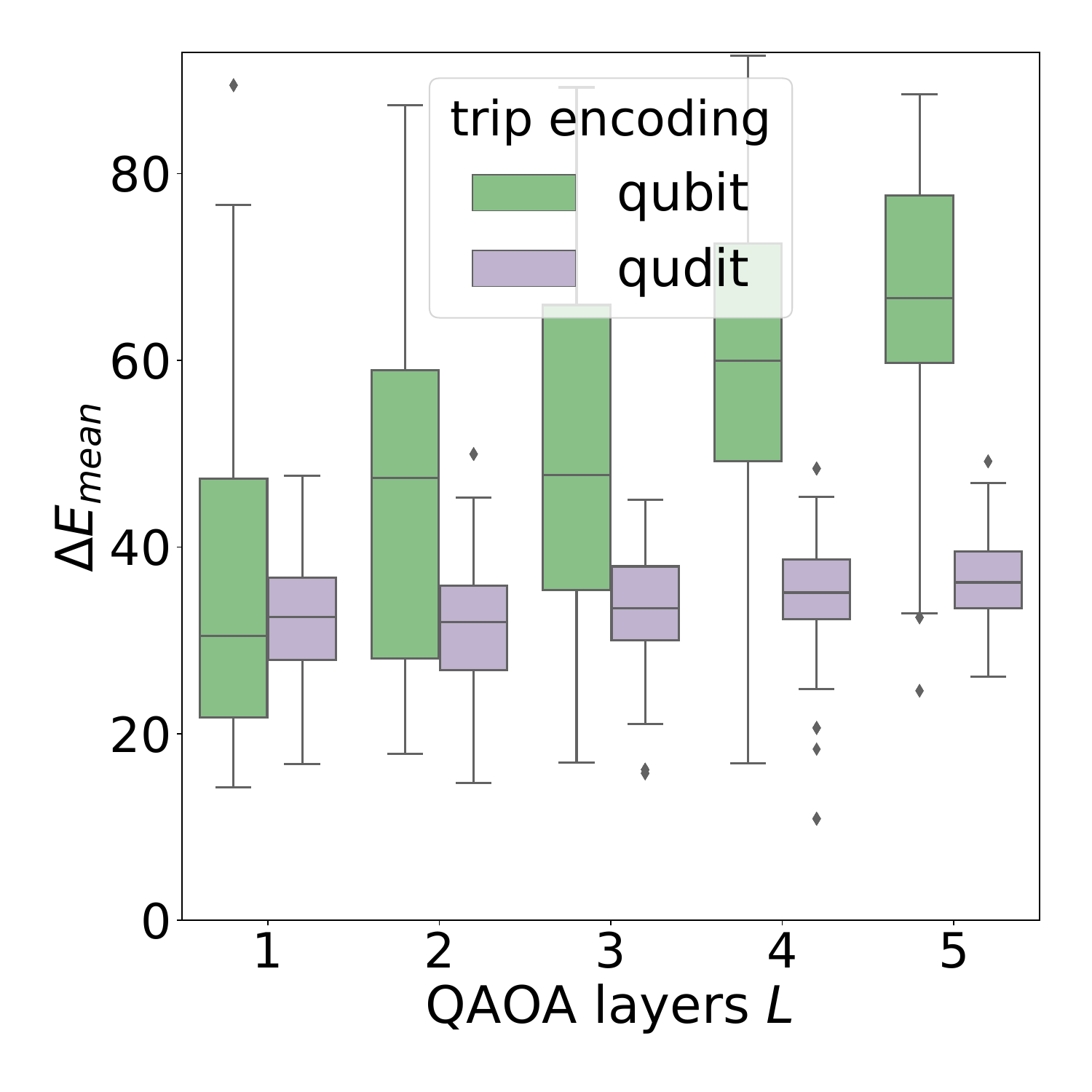}    \includegraphics[width=0.24\linewidth]{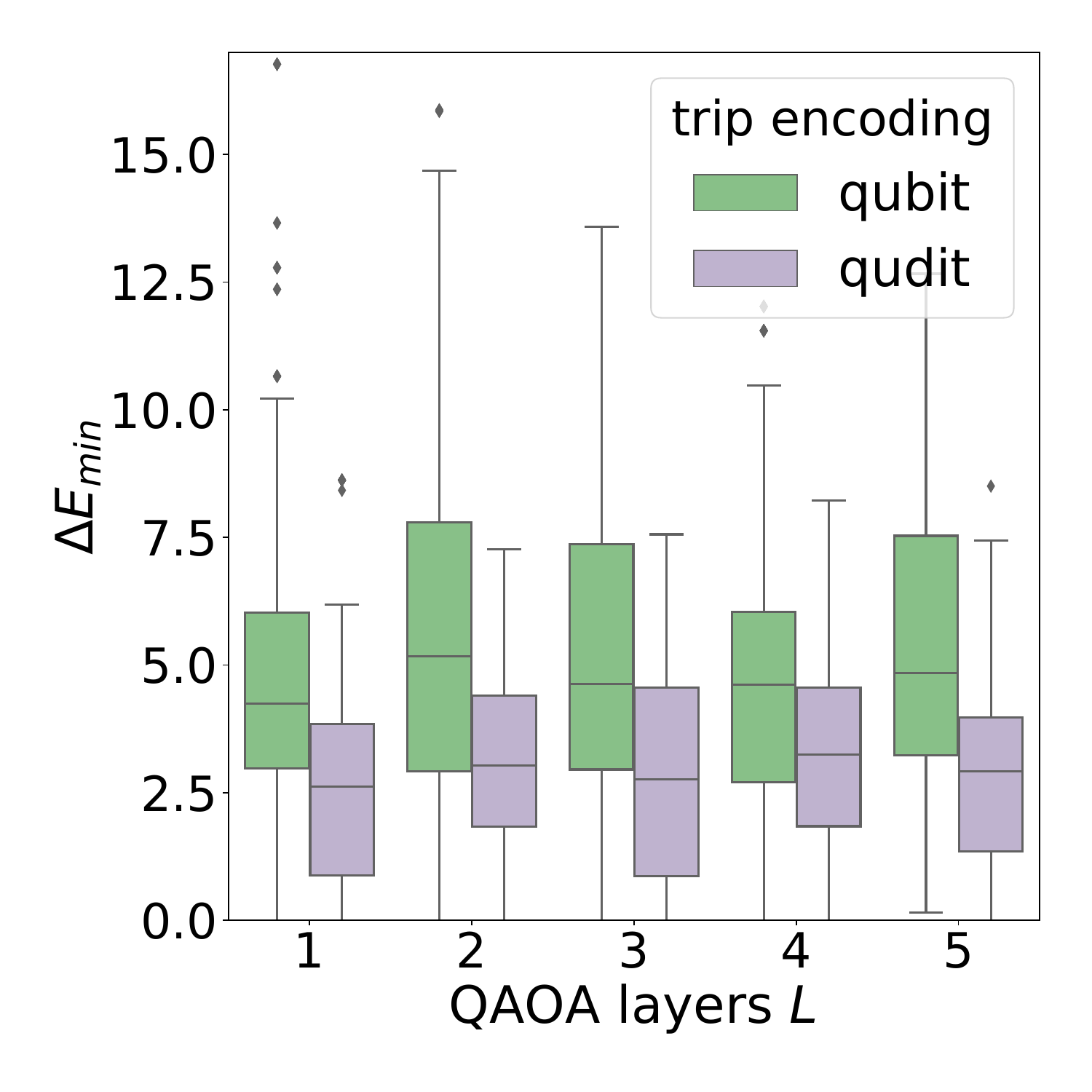}
    \includegraphics[width=0.24\linewidth]{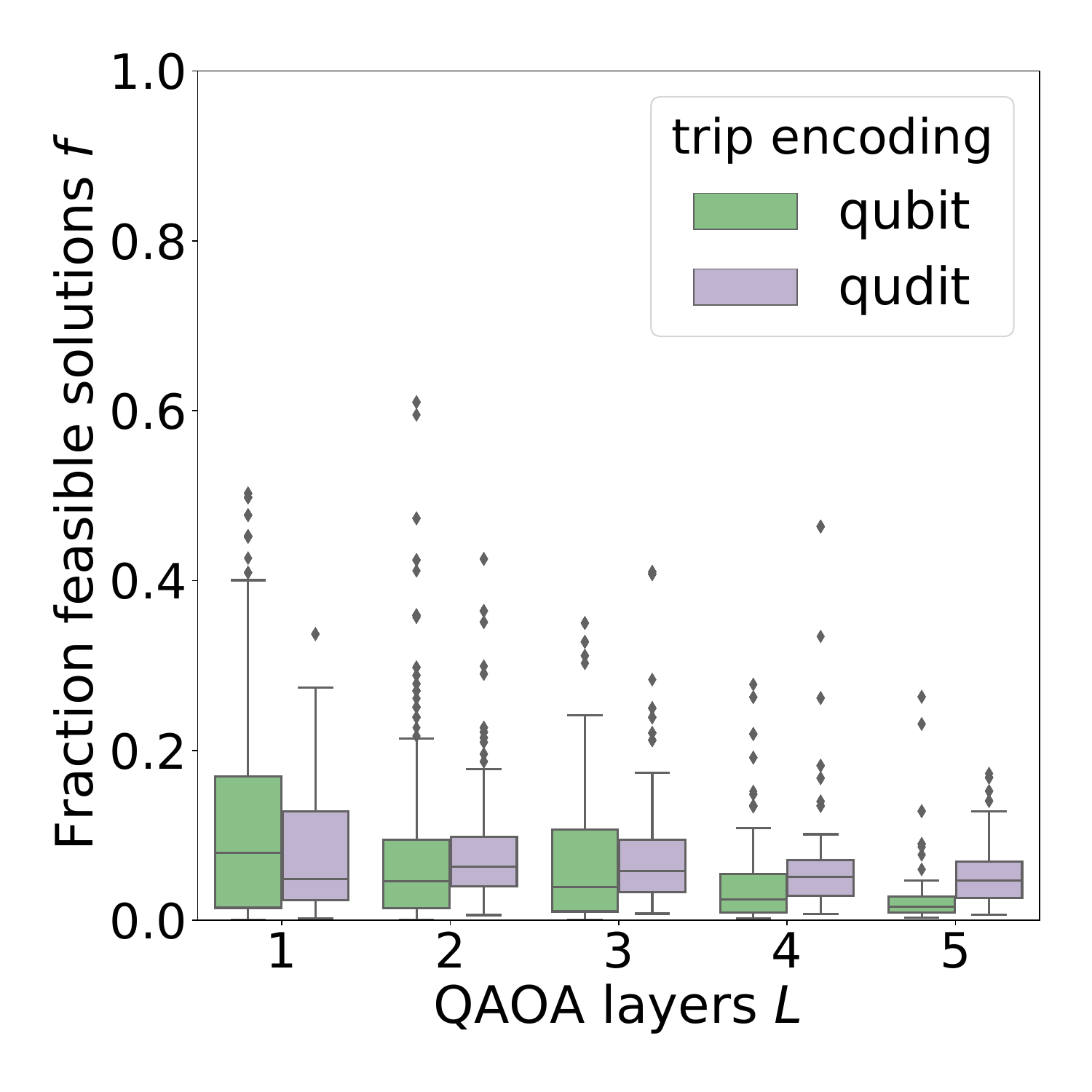}
    \includegraphics[width=0.24\linewidth]{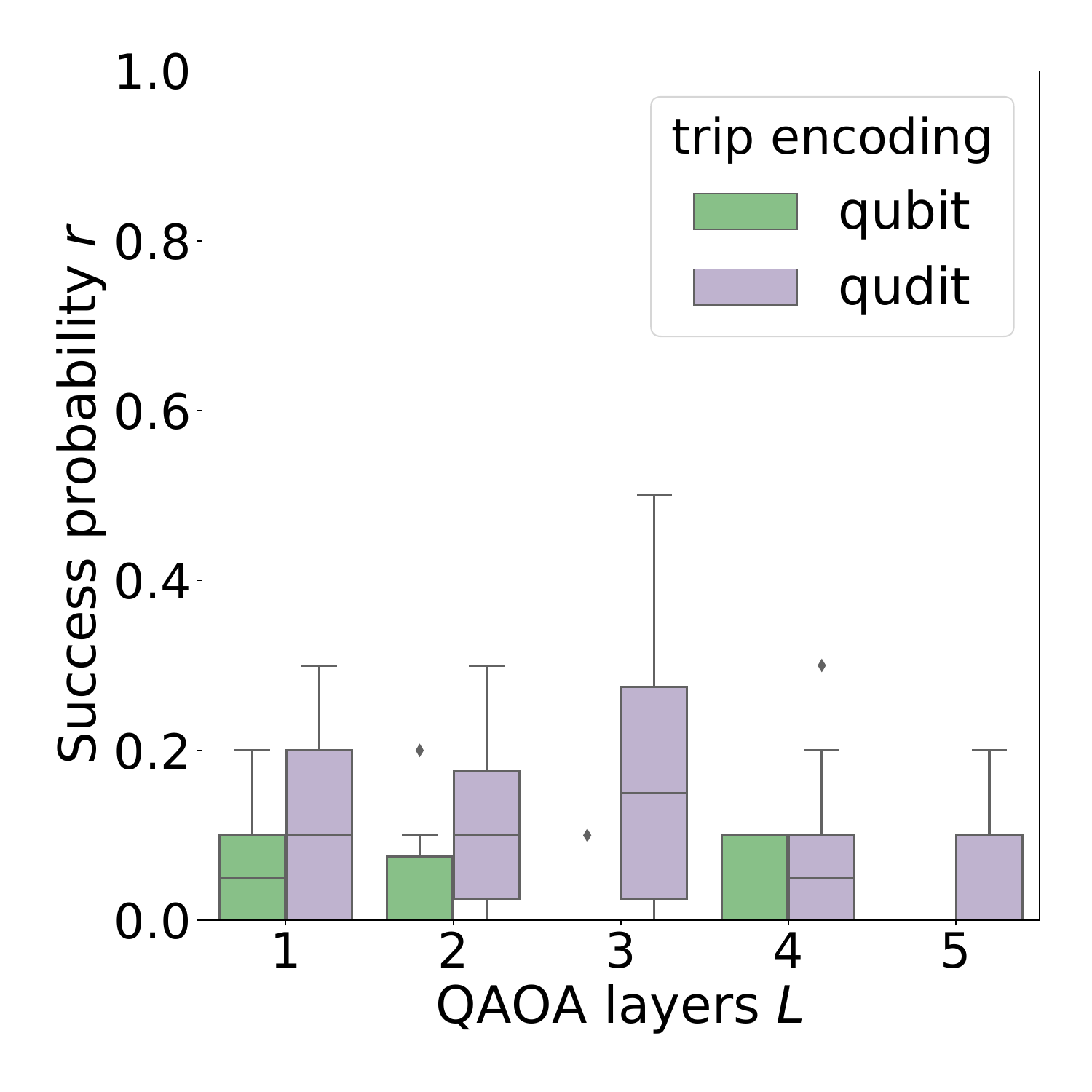}
    \caption{QAOA performance for qudit vs qubit trip encodings averaged over ten problem instances for the bi-directional charging ($d=3$) optimization with $N_{EV}=3$, $T=2$, and $R=2$ and $N_\text{runs}=10$ runs each as a function of circuit depth $L$. 
    From left to right: mean energy gap $\Delta E_{mean}$ of \Cref{eq:ev_Emean}, minimal energy gap $\Delta E_{min}$ of \Cref{eq:ev_Emin}, feasible-sample fraction $f$ of \Cref{eq:ev_feasible}, and success probability $r$ of \Cref{eq:ev_success}. 
    All runs use $200$ Powell iterations and use $M=256$ shots for calculating the expectation values. 
    The qudit encoding produces slightly better optimization results with lower variance in energy and more robust performance for large $L$.     
    }
    \label{fig:ev_trio_results}
\end{figure*}

\Cref{fig:ev_trio_results} summarizes the results of qubit and qudit trip encodings for the bi-directional charging optimization with three charging levels ($d=3$) and for $N_{EV}=3$ EVs, $T=2$ time steps, and $R=2$ customer trip reservations as a function of QAOA depth $L$.

The mean energy gap of \Cref{eq:ev_Emean} is shown in the first (leftmost) panel. 
Overall, the absolute values as well as the variance of the mean energy are larger for the qubit encoding compared to the qudit case. 
In addition, the values are constant as a function of QAOA layers for the qudit encoding, while they increase substantially with increasing $L$ for the qubit encoding.  
The second panel (middle left) shows the minimal energy gap of \Cref{eq:ev_Emin} extracted from the optimized quantum circuit. 
Similar to $\Delta E_{mean}$, the qudit representation produces slightly better results with lower variance than the qubit encoding. 
However, the increase in energy with increasing layers for the qubit encoding is not as pronounced.  
The third (middle right) panel shows the fraction of feasible solutions $f$ of \Cref{eq:ev_feasible}, i.e.\ the number of states which fulfill all constraints, and the last (rightmost) panel shows the success probability $r$ of \Cref{eq:ev_success}, both sampled from the final quantum circuit after the optimization. 
Both plots show that the optimization substantially improves the probabilities of obtaining feasible and optimal states compared to the bare probabilities shown in \Cref{fig:problem_stats}. 
However, the fraction of feasible states is only on the order of $\sim10\%$ which is still rather low. 
This leads to the large mean energies observed in the leftmost panel since many of the sampled solutions are infeasible and acquire a large contribution from the penalty terms. 

\begin{figure}
    \centering
    \includegraphics[width=0.98\linewidth]{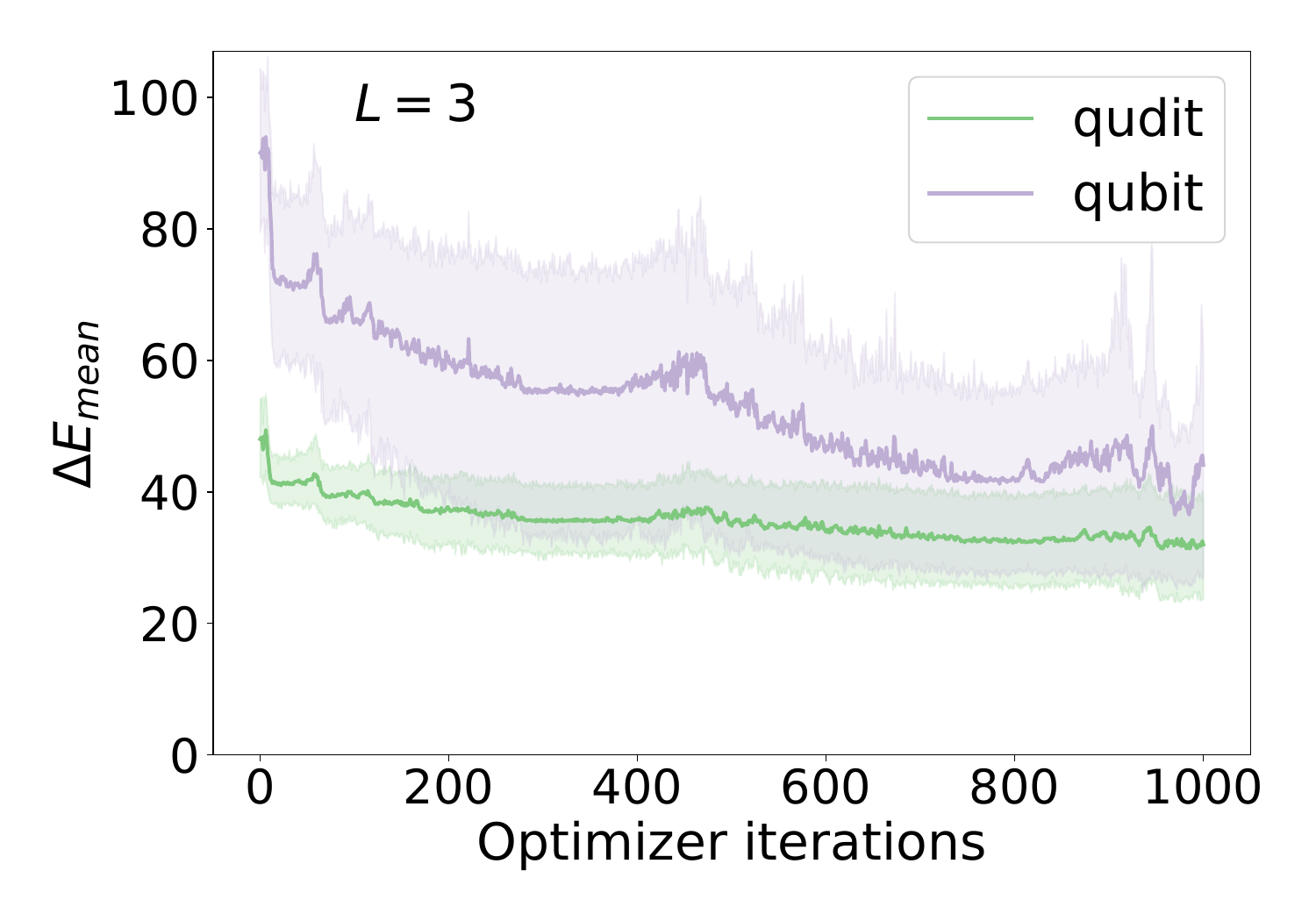}
    \caption{QAOA optimization progress performance for qudit and qubit trip encodings over ten instances of the bi-directional ($d=3$) charging problem with $N_{EV}=3$, $T=2$, and $R=2$ averaged over $N_\text{runs}=10$ runs each for fixed depth $L=3$. All runs use $1000$ Powell iterations and $M=256$ shots. 
    The reduction in Hilbert space dimension for the qudit encoding results in quicker convergence. 
    }
\label{fig:ev_trio_progress}
\end{figure}

The overall picture of these results is that the qudit encoding yields slightly better results, with much tighter variance, than the qubit encoding. 
Surprisingly, the results do not improve with increasing QAOA layers, which would be expected given the greater expressivity of the variational circuit with more layers and therefore the increased likelihood of producing optimized quantum circuits.  
The observed behavior can be explained by the increasing number of variational parameters with depth (here, $\sim 3L$), which makes the optimization problems harder as $L$ increases.
This typically leads to slower convergence for more variational parameters. 
However, because we limit the number of optimization iterations to 200 regardless of the number of parameters, we obtain increasingly nonconverged solutions as the number of layers increases. 
In the case of qudit encoding, the potential performance gain from greater circuit expressivity is apparently offset by the premature termination of the optimization. 
For the qubit encoding with larger Hilbert space, the optimization progress is much slower, and the impact of additional search variables and premature stopping of the optimization even leads to the observed decrease in performance.  
This can be explicitly observed in \Cref{fig:ev_trio_progress}, which shows the optimization progress of the Powell optimizer for a QAOA with a fixed number of layers ($L=3$) up to 1000 iterations.
Both encodings exhibit a sharp decrease in the first couple of iterations, followed by a slight decrease as the number of iterations increases. 
While both runs seem to converge to the same low-energy solutions, the progress is much slower for the qubit encoding.
Therefore, stopping at 200 iterations has a much more severe impact on the performance for the qubit encoding than the qudit encoding.

\begin{figure}
    \centering
\includegraphics[width=0.84\linewidth]{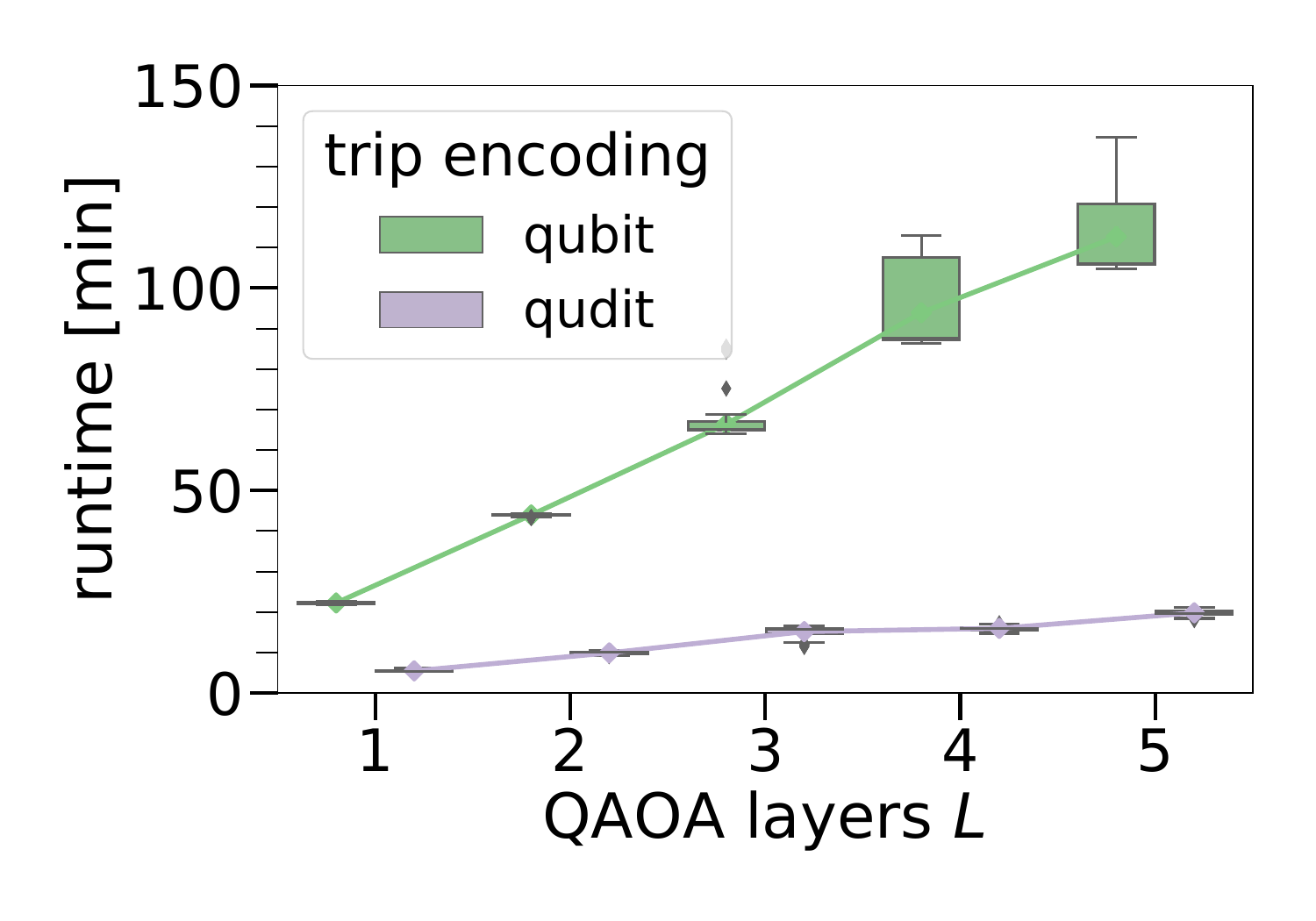}
    \caption{Mean runtime for one Powell optimization run with $200$ iterations and $M=256$ shots  for the bi-directional charging ($d=3$) with $N_{EV}=3$, $T=2$ and $R=2$  versus  circuit depth $L$. 
    Simulations utilizing the qudit encoding require substantially shorter runtimes.     
    }
    \label{fig:ev_runtime}
\end{figure}

\Cref{fig:ev_runtime} shows the statistics of the simulation runtime in minutes for one optimization with 200 Powell iterations.
The simulation runtime increases linearly with the QAOA circuit depth, as expected, since the number of matrix operations scales linearly with $L$. 
A large runtime difference is observed between the qubit and qudit encoding, which is related to the exponential reduction in Hilbert space dimension from qubits to qudits.
The state-vector simulation runtime is not directly relevant for the expected runtime on real quantum hardware.
However, a more compact representation can reduce the overhead of implementing constraints and logical variables, which in turn can translate into fewer single- and two-qudit gate operations.
Of course, these effects are hardware dependent, and evaluating them in detail requires an explicit circuit-level implementation of the qubit and qudit circuits. 
Therefore, the simulation runtime can be viewed as a proxy for representational complexity; implementing more efficient encodings might be beneficial for classical prototyping as well as implementations on quantum hardware.

\subsection{Uni-directional charging}
\label{sec:numerical-results_d2}
We now focus on ten problem instances of the uni-directional ($d=2$) charging optimization with $N_{EV}=2$, $T=3$ and where we vary the number of trips $R=\{2,3,4\}$.   
Because we only consider two charging levels, we have a binary optimization variable for the charging part of the Hamiltonian for both encodings. 
This implies pure qubit systems for the qubit encoding while the qudit encoding is still a heterogeneous mix of qubits and qudits. Like previously mentioned at the end of \Cref{sec:quantum-optimization-algorithm}, this means that the qubit encoding has fewer variational parameters as a function of layers compared with the qudit case. 
Therefore, all the optimization runs below have this structural advantage for the qubit encoding. 
As we will see, even with this advantage, the qudit encoding still outperforms similarly to the bi-directional charging problem.

\begin{figure}
    \centering
    \includegraphics[width=0.48\linewidth]{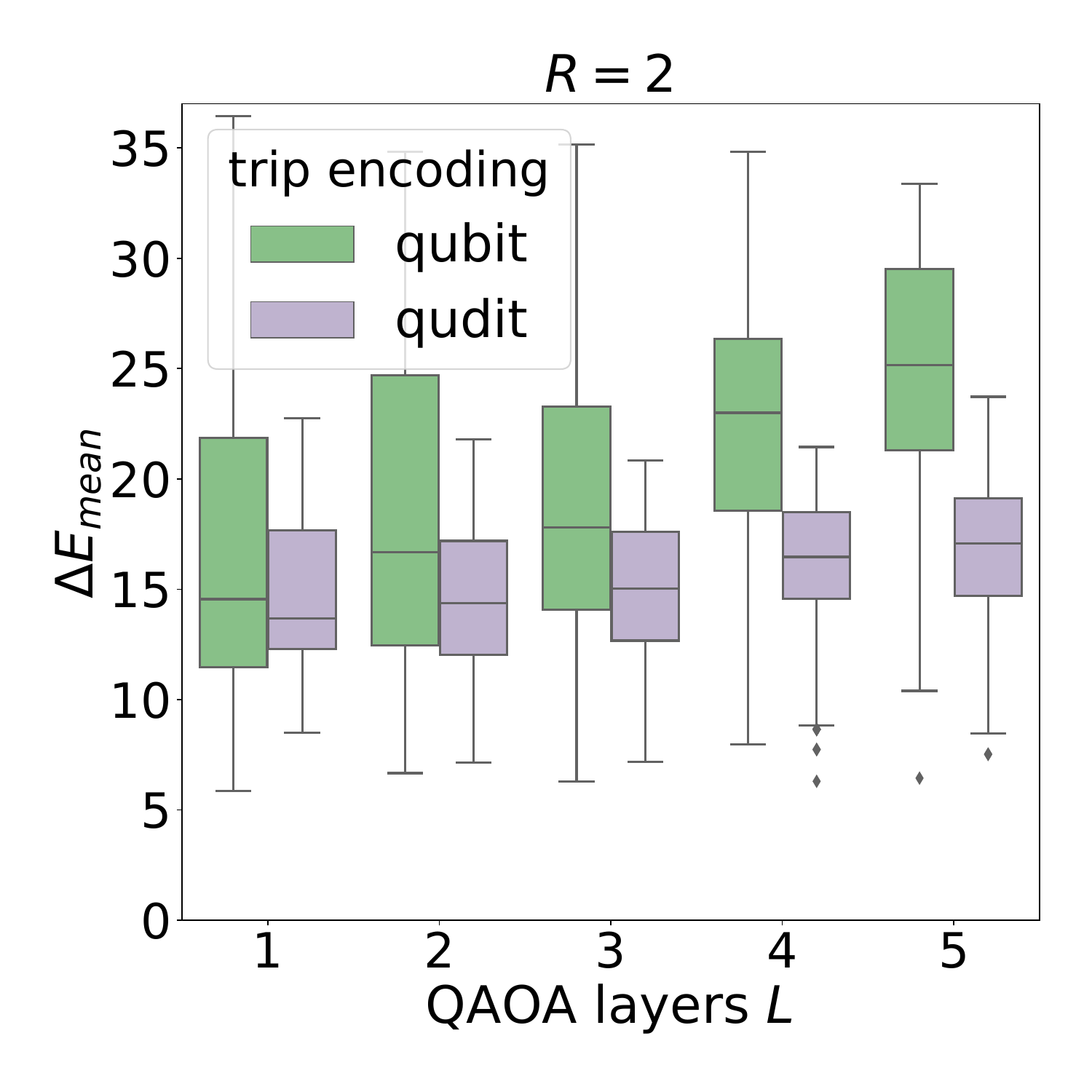}
    \includegraphics[width=0.48\linewidth]{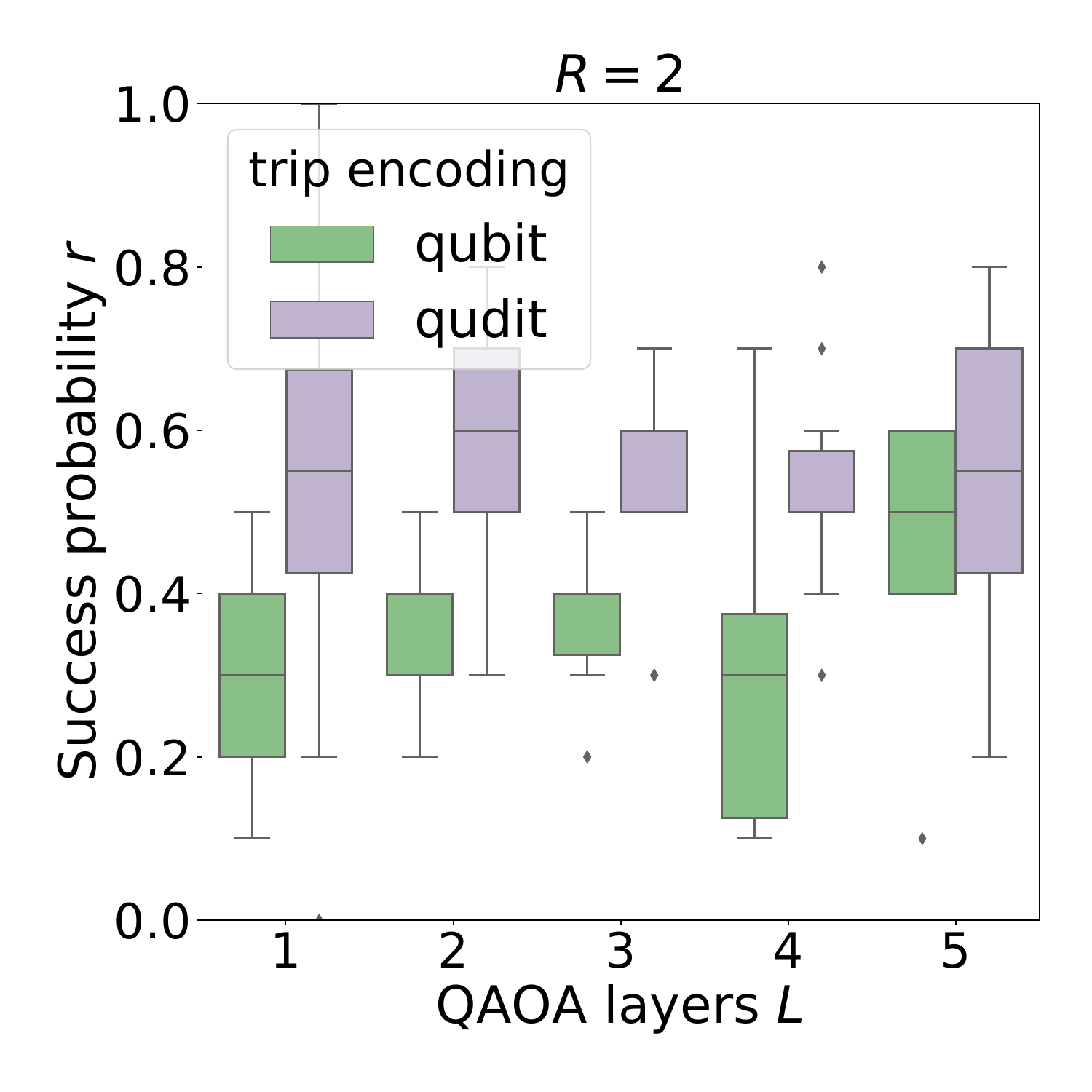}
\\
   \includegraphics[width=0.48\linewidth]{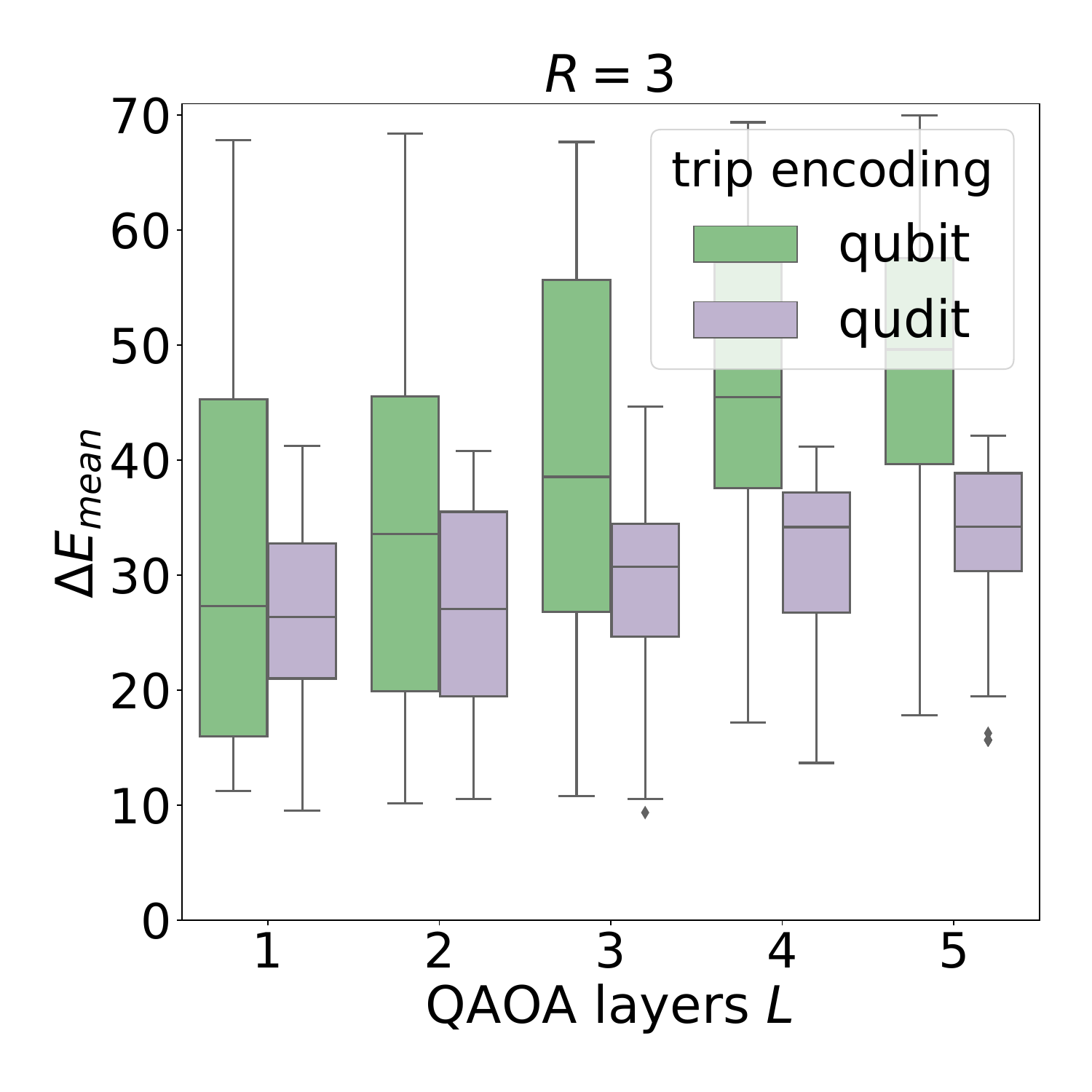}
    \includegraphics[width=0.48\linewidth]{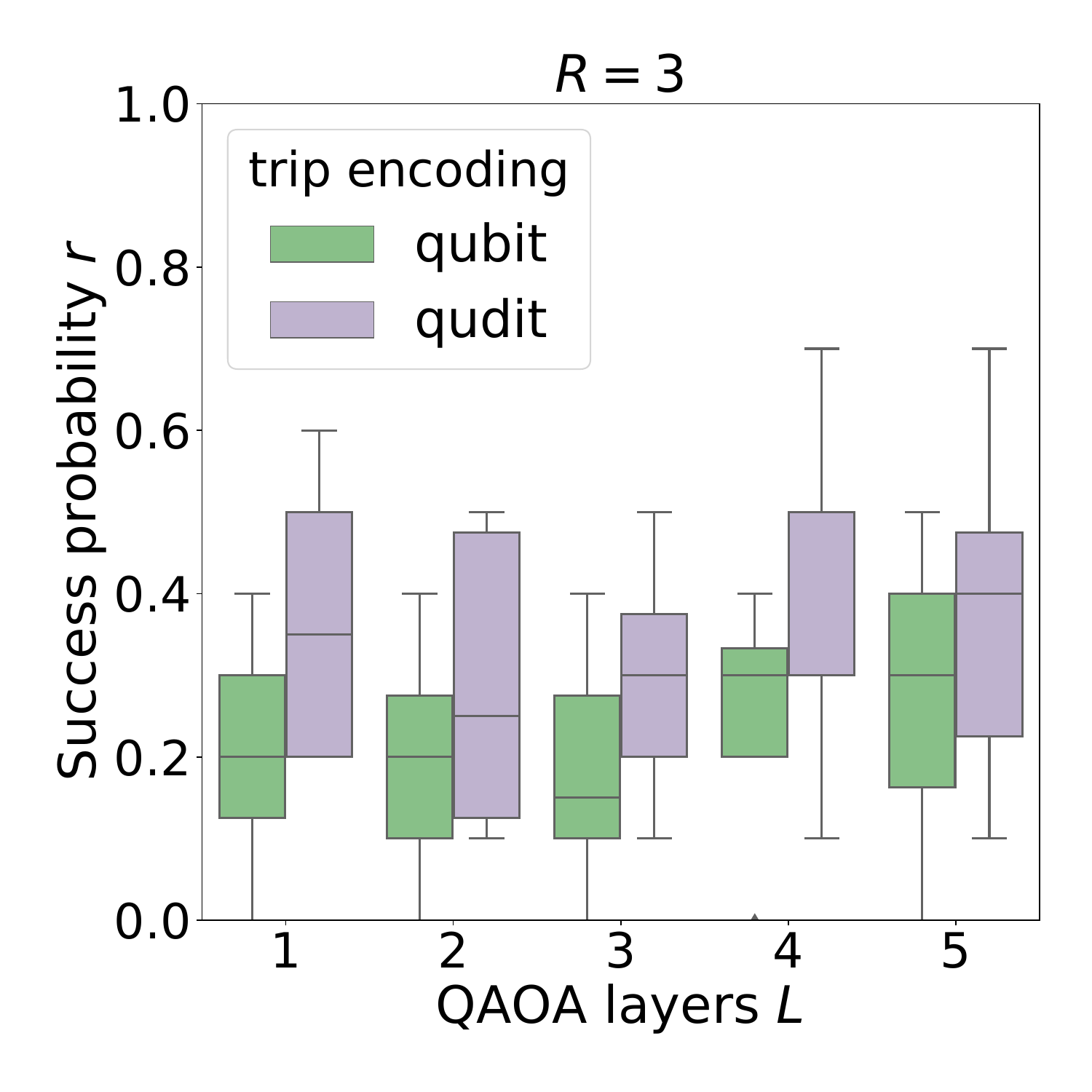}
   \\
    \includegraphics[width=0.48\linewidth]{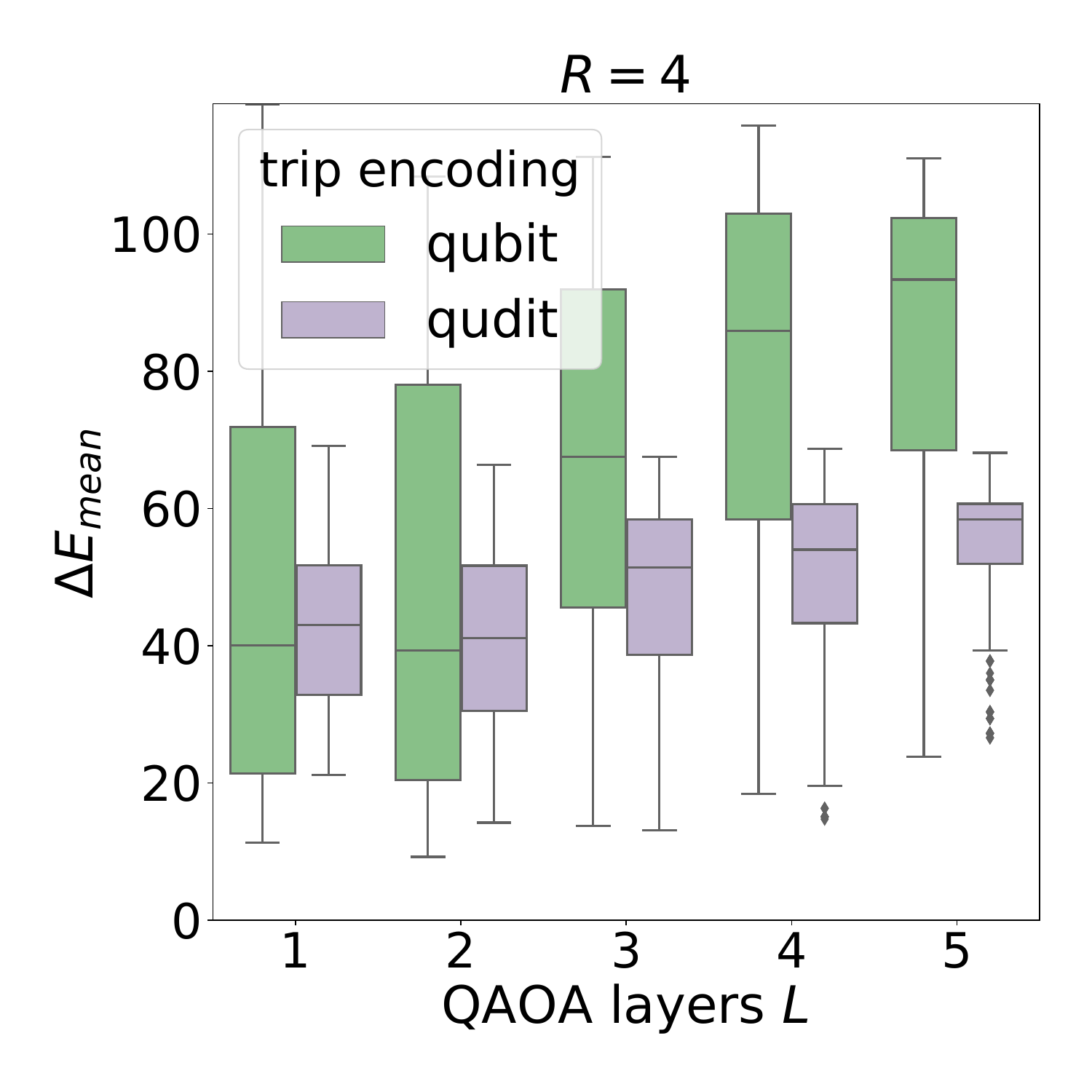}
    \includegraphics[width=0.48\linewidth]{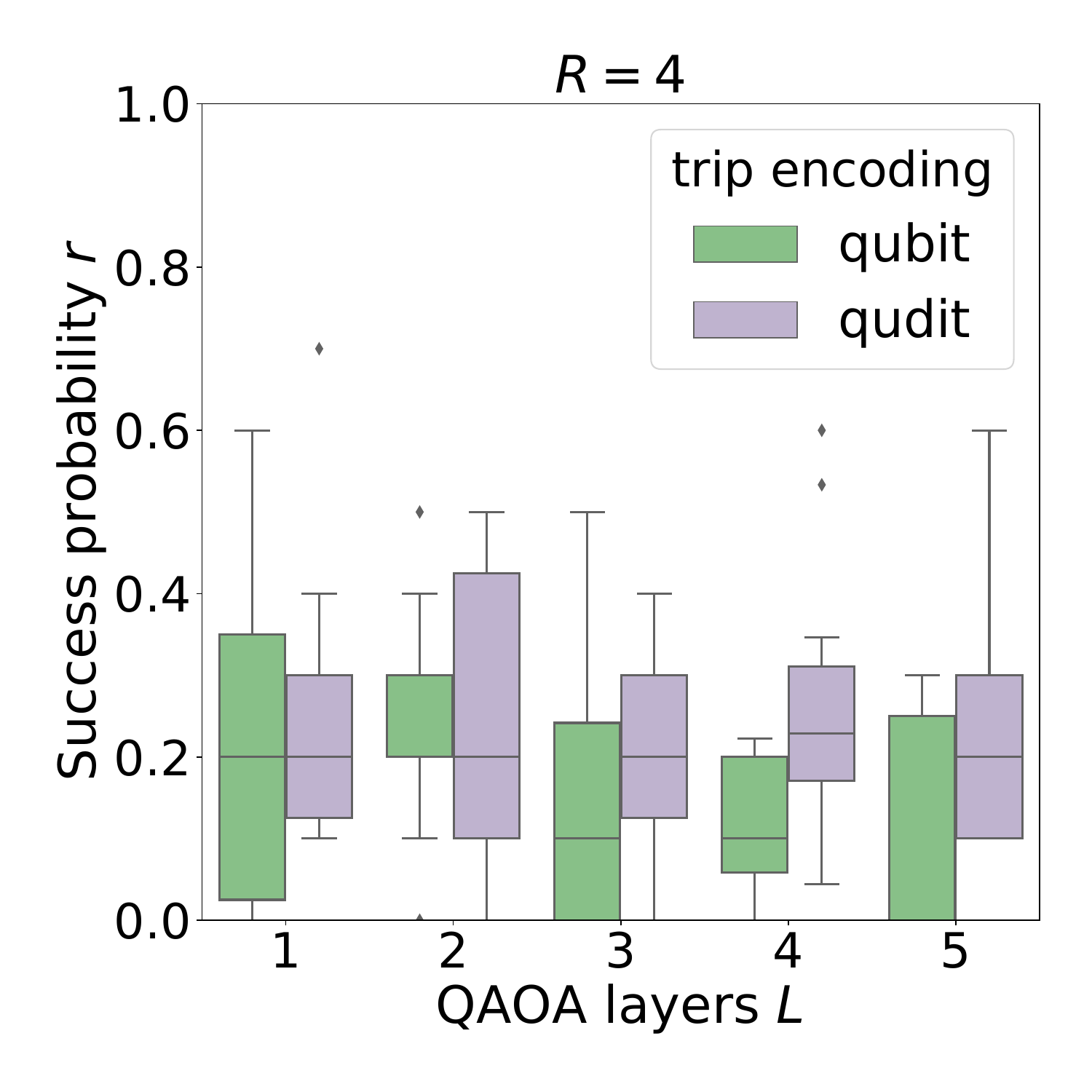}
    \caption{QAOA performance as a function of layers $L$ for qudit vs qubit trip encodings over ten instances and $N_\text{runs}=10$ runs each for uni-directional ($d=2$) charging problem instances with $N_{EV}=2$, $T=3$ and for $R=2$ (upper row),  $R=3$ (middle row) and  $R=4$ (lower row) trip reservations. 
    The left column shows the mean energy gap $\Delta E_{\mathrm{mean}}$ of \Cref{eq:ev_Emean} after the optimization, while the right column shows the success probabilities $r$ of \Cref{eq:ev_success}. 
    All runs use $200$ Powell iterations and $M=256$ shots.
    The qudit encoding produces slightly better optimization results.   
    }
    \label{fig:ev_trip_results_d2}
\end{figure}

\Cref{fig:ev_trip_results_d2} compares qubit and qudit trip encodings as a function of QAOA depth $L$.
The qubit encoding yields higher mean energies, lower success probabilities, and greater variance in the results.  
Also, the performance decrease of the qubit encoding with increasing circuit depth, due to limited optimization iterations, can be observed.
In contrast, the qudit encoding is again more robust and almost maintains the same performance for all layers, even though there are more parameters to optimize for the qudit encoding. 
The results also show that the problems get increasingly more constrained for larger $R$, and consequently, the mean energies increase while the success probabilities decrease for larger $R$.

\section{Conclusion}
\label{sec:conclusion}
We investigated QAOA for a constrained EV charging and trip assignment problem, comparing binary and integer trip encodings by direct Hamiltonian simulation.
Both encodings describe the same feasible schedules and only differ in encodings of infeasible configurations. 
The qubit encoding uses several qubits to represent which trips are served by which EV, where additional constraints must be imposed to ensure the feasibility of the encoding. 
In contrast, the qudit encoding uses one high-dimensional qudit for each reservation to encode the assigned EV, which reduces the required degrees of freedom and eliminates many encoding constraints. This yields an exponential reduction in Hilbert-space dimension with the number of trips, see \Cref{eq:hilbertDim}, as well as reducing the overhead of handling constraints.

We conducted extensive numerical simulations of many different random instances of such realistic problems. Our benchmarks operated in a highly constrained regime, in which only a small fraction of possible configurations are feasible. In this setting, the encoding choice directly affects the probability assigned to infeasible configurations. The qudit encoding reduces the total configuration space while preserving the feasible set, which is consistent with the higher feasible-sample fractions and success rates we observe at comparable depth.

In our experiments, the qudit encoding consistently achieved optimization performance comparable to, or slightly better than, that of the more conventional qubit encoding. Furthermore, across all experiments, the variance of mean energies of the qudit encoding results was  smaller than for the qubit case. 
At the same time, the required computational resources measured by Hilbert space dimensions and simulation runtime are substantially reduced for the qudit encoding. 
With a fixed budget on the classical optimization iterations, a degradation of the performance is expected with increasing circuit complexity (QAOA layers), which is due to the slower optimization convergence for more search variables.
Our work demonstrates that the efficiency of the qudit encoding is highly advantageous in such a constrained setting, with almost no performance degradation observed. 
The optimization proceeds much faster than for the qubit encoding, and therefore, the qubit encoding suffers much more strongly from the premature stopping of the optimization than the qudit encoding.    
This suggests that smaller, more structured encodings make the variational search problem more robust in realistic settings with limited computational resources.

An important next step is to test whether these trends are specific to this EV benchmark or reflect a more general effect of encoding choice in constrained quantum optimization. This motivates scaling to larger instances and studying a broader set of industrially relevant constrained optimization problems with sparse feasibility, to understand how encoding interacts with QAOA depth, shot noise, and a fixed classical optimization budget. It would also be valuable to move toward hardware-oriented implementations, where deeper circuits and explicit constraint handling (e.g.,~\cite{bottarelliConstraints2024}) amplify resource costs and expose noise sensitivity, and to compare qubit and qudit realizations on quantum hardware to see how much of the observed advantage persists beyond state-vector simulation.

\section*{Acknowledgements}
The authors acknowledge fruitful discussions with colleagues at the Honda Research Institute Europe, and extend their gratitude to X.~Bonet-Monroig, A.~Bottarelli, V.~Dunkjo, P.~Hauke, M.~Olhofer, J.~Schmüdderich, H.~Wersing, and B.~Sendhoff.
LE and SS acknowledge funding from the NeQST project of the European Union under Horizon European Program, Grant Agreement 101080086.
Views and opinions expressed are those of the author(s) only and do not necessarily reflect those of the European Commission. 
Neither the European Union nor the granting authority can be held responsible for them.

\appendix

\section{Angular Momentum Operators}
For the interested reader, we provide details here on the angular momentum operators that represent the charging and trip-assignment variables in the optimization problem. 
We show the expressions for  bi-directional charging. The uni-directional case is  essentially the same but with a straightforward shift  in the eigenvalues $l_{n,t}$. 
\label{sec:ang-momentum}
\subsection{Qubit encoding}
For the binary trip encoding, the computational basis is indexed by qudit charging variables $l_{n,t}$ and qubit trip-assignment variables $r_{n,i}$, i.e., basis states of the form $\ket{l_{n,t},r_{n,i}}$. 
The corresponding angular momentum operators are 
\begin{align}
L^z_{n,t} \ket{l_{n,t},r_{n,i}} & = l_{n,t}\ket{l_{n,t},r_{n,i}},\\  
L^{+}_{n,t} \ket{l_{n,t},r_{n,i}}   &= \sqrt{(\tfrac{d-1}2-l_{n,t})(\tfrac{d+1}2+l_{n,t})}\ket{l_{n,t}+1,r_{n,i}}  , \\ 
L^{-}_{n,t} \ket{l_{n,t},r_{n,i}}  &= \sqrt{(\tfrac{d-1}2+l_{n,t})(\tfrac{d+1}2-l_{n,t})} \ket{l_{n,t}-1,r_{n,i}}, \\
R^z_{n,i} \ket{l_{n,t},r_{n,i}} & = r_{n,i}\ket{l_{n,t},r_{n,i}},\\
R^{+}_{n,i} \ket{l_{n,t},r_{n,i}}   &= \sqrt{(1-r_{n,i})(1+r_{n,i})}\ket{l_{n,t},r_{n,i}+1}  ,\\
R^{-}_{n,i} \ket{l_{n,t},r_{n,i}}  &= \sqrt{r_{n,i}\,(2-r_{n,i})} \ket{l_{n,t},r_{n,i}-1}
\end{align}
where the eigenvalues correspond to charging power levels and trip assignment variables, 
$l_{n,t}=\{-\tfrac{d-1}2,\dots,0,\dots,\tfrac{d-1}2\}$ ($d$ odd) and $r_{n,i}=\{0,1\}$
and 
\begin{align}
    \label{eq:defLx}
    L^x_{n,t} &= \frac{1}{2} (L^+_{n,t} + L^-_{n,t}) \, ,& R^x_{n,i} &= \frac{1}{2} (R^+_{n,i} + R^-_{n,i})   \\
    L^y_{n,t} &= \frac{1}{2i} (L^+_{n,t} - L^-_{n,t}) \, , &   R^y_{n,i} &= \frac{1}{2} (R^+_{n,i} - R^-_{n,i}) . 
\end{align}

\subsection{Qudit encoding}
For the integer encoding of the trip variables, basis states are indexed by the same qudit variables $l_{n,t}$ as before, while the trip assignment is encoded with qudit variables $q_{i}$, 
\begin{align}
L^z_{n,t}\ket{l_{n,t},q_i} &= l_{n,t}\ket{l_{n,t},q_i},
\\
Q^z_{i}\ket{l_{n,t},q_i} &= q_i\ket{l_{n,t},q_i},
,
\end{align}
where $ q_i\in\{0,\dots,N_{EV}\}$.
We use the standard angular-momentum operators for both qudit types, with $L^{x,y}_{n,t}$ and $Q^{x,y}_i$ defined in the usual way from the corresponding ladder operators,
\begin{align}
L^z_{n,t} \ket{l_{n,t},q_{i}} & = l_{n,t}\ket{l_{n,t},q_{i}},\\
L^{+}_{n,t} \ket{l_{n,t},q_{i}}   &= \sqrt{(\tfrac{d-1}2-l_{n,t})(\tfrac{d+1}2+l_{n,t})}\ket{l_{n,t}+1,q_{i}}  ,\\ 
L^{-}_{n,t} \ket{l_{n,t},q_{i}}  &= \sqrt{(\tfrac{d-1}2+l_{n,t})(\tfrac{d+1}2-l_{n,t})} \ket{l_{n,t}-1,q_{i}}, \\ 
Q^z_{i} \ket{l_{n,t},q_{i}} & = q_{i}\ket{l_{n,t},q_{i}},\\
Q^{+}_{i} \ket{l_{n,t},q_{i}}   &= \sqrt{(N_{EV}-q_{i})(q_{i}+1)}\ket{l_{n,t},q_{i}+1}  ,\\ 
Q^{-}_{i} \ket{l_{n,t},q_{i}}  &= \sqrt{q_{i}(N_{EV}+1-q_{i})} \ket{l_{n,t},q_{i}-1}, \\ 
Q^{+}_{n,i} \ket{l_{n,t},q_{i}}   &= \sqrt{(1-q_{i})(1+q_{i})}\ket{l_{n,t},q_{i}+1}  \\
Q^{-}_{n,i} \ket{l_{n,t},q_{i}}  &= \sqrt{q_{i}\,(2-q_{i})} \ket{l_{n,t},q_{i}-1}
\end{align}
where the eigenvalues correspond to charging power levels and trip assignment variables.

\bibliographystyle{splncs04}
\bibliography{ref.bib}

\end{document}